\documentclass[conference]{IEEEtran}
\IEEEoverridecommandlockouts
\usepackage{cite}
\usepackage{amsmath,amssymb,amsfonts}
\usepackage{algorithmic}
\usepackage{graphicx}
\usepackage{textcomp}
\usepackage{xcolor}
\usepackage{framed}
\usepackage{booktabs}
\usepackage{multirow}
\usepackage{balance}
\usepackage{float}
\usepackage{tikz}
\def\BibTeX{{\rm B\kern-.05em{\sc i\kern-.025em b}\kern-.08em
    T\kern-.1667em\lower.7ex\hbox{E}\kern-.125emX}}
\begin{document}

\newcommand\copyrighttext{%
      \footnotesize \textcopyright~2025 IEEE. This document is a preprint. Personal use of this material is permitted.
      Permission from IEEE must be obtained for all other uses, in any current or future
      media, including reprinting/republishing this material for advertising or promotional
      purposes, creating new collective works, for resale or redistribution to servers or
      lists, or reuse of any copyrighted component of this work in other works. DOI: tbd
      }
    \newcommand\copyrightnotice{%
    \begin{tikzpicture}[remember picture,overlay]
    \node[anchor=north,yshift=-12pt] at (current page.north) {\fbox{\parbox{\dimexpr\textwidth-\fboxsep-\fboxrule\relax}{\copyrighttext}}};
    \end{tikzpicture}%
}

\title{Identifying Explanation Needs: Towards a Catalog of User-based Indicators 
}

\makeatletter 
\newcommand{\linebreakand}{%
  \end{@IEEEauthorhalign}
  \hfill\mbox{}\par
  \mbox{}\hfill\begin{@IEEEauthorhalign}
}
\makeatother 

\author{\IEEEauthorblockN{Hannah Deters}
\IEEEauthorblockA{\textit{Software Engineering Group} \\
\textit{Leibniz University Hannover}\\
Hannover, Germany \\
hannah.deters@inf.uni-hannover.de}
\and
\IEEEauthorblockN{Laura Reinhardt}
\IEEEauthorblockA{\textit{Software Engineering Group} \\
\textit{Leibniz University Hannover}\\
Hannover, Germany \\
laura.reinhardt@inf.uni-hannover.de}
\and
\IEEEauthorblockN{Jakob Droste}
\IEEEauthorblockA{\textit{Software Engineering Group} \\
\textit{Leibniz University Hannover}\\
Hannover, Germany \\
jakob.droste@inf.uni-hannover.de}
\linebreakand
\IEEEauthorblockN{Martin Obaidi}
\IEEEauthorblockA{\textit{Software Engineering Group} \\
\textit{Leibniz University Hannover}\\
Hannover, Germany \\
martin.obaidi@inf.uni-hannover.de}
\and
\IEEEauthorblockN{Kurt Schneider}
\IEEEauthorblockA{\textit{Software Engineering Group} \\
\textit{Leibniz University Hannover}\\
Hannover, Germany \\
kurt.schneider@inf.uni-hannover.de}
}

\maketitle

\copyrightnotice
\vspace{-2ex}

\begin{abstract}
In today's digitalized world, where software systems are becoming increasingly ubiquitous and complex, the quality aspect of explainability is gaining relevance. 
A major challenge in achieving adequate explanations is the elicitation of individual explanation needs, as it may be subject to severe hypothetical or confirmation biases. 
To address these challenges, we aim to establish user-based indicators concerning user behavior or system events that can be captured at runtime to determine when a need for explanations arises. 
In this work, we conducted an online study to collect self-reported indicators that could indicate a need for explanation. We compiled a catalog containing 17 relevant indicators concerning user behavior, 8 indicators concerning system events and 14 indicators concerning emotional states or physical reactions. We also analyze the relationships between these indicators and different types of need for explanation. 
The established indicators can be used in the elicitation process through prototypes, as well as after publication to gather requirements from already deployed applications using telemetry and usage data.
Moreover, these indicators can be used to trigger explanations at appropriate moments during the runtime. 
\end{abstract}

\begin{IEEEkeywords}
explainability, requirements engineering, user experience
\end{IEEEkeywords}

\section{Introduction}
Explainability is a quality aspect that describes the ability of a software system to explain itself~\cite{chazette2021exploring}. This includes explaining various system aspects such as system behavior~\cite{adadi2018peeking,habiba2025ml}, interactions with the system~\cite{droste2024explanations, Deters2023explanationsOnDemand}, or privacy~\cite{brunotte2023context,brunotte2023privacy}. 
One challenge in the area of explainability is the highly individual needs of users~\cite{Droste2023personas, ramos2021modeling}. Depending on the user and the context of use, different explanations may be needed~\cite{chazette2021exploring}. For example, a user's prior knowledge can influence whether an explanation is needed at all~\cite{schneider2019personalizedexplanationmachinelearning}, and personal preferences can influence which type of presentation is preferred~\cite{id196_MILLER20191}. Additionally, the same user may require different explanations in different contexts~\cite{chazette2021exploring}. In calm settings, users can process longer, more in-depth explanations, whereas in stressful, hectic contexts, only very brief, essential explanations should be provided. This challenge is further compounded the fact that explanations should only be shown when they are actually needed, as they can also have negative effects. For instance, explanations can be stressful for the user~\cite{deters2024x, gruning2024stressful} and increase their mental load~\cite{nunes2017systematic}. Explanations can also reduce trust in the system~\cite{Kizilcec2016howMuchInformation} if they reveal flaws in what users perceive as a flawless system. Therefore, the correct identification of the need for explanations is very important in order to provide only desired explanations. 

However, eliciting the need for explanations has proven to be challenging~\cite{droste2024explanations}. When users are simply asked whether they want a particular explanation, they tend to accept the offer since it appears to come at no cost. Droste et al.~\cite{Droste2023personas} referred to this phenomenon as the \textit{why-not mentality}. In contrast, providing open-ended questions instead of concrete explanation examples  avoids the \textit{why-not mentality} but introduces a so-called hypothetical bias~\cite{hypotheticalBiasPlott}, since users have to think about which explanations they might have wanted. Therefore, asking users about their need for explanations is difficult to do without introducing bias.
To provide an alternative that prevents these biases, we aim to establish user-based indicators that can be recorded at runtime and that objectively indicate a need for explanation. These indicators include specific user behaviors within in the system (e.g., back-and-forth navigation) or physical reactions of the users (e.g., facial expressions). 

To identify these indicators, we conducted an online study with 66 participants. Participants reported their needs for explanation for three software systems they had recently used. They then reported what they did and how they felt when they encountered the uncertainty that led to a need. These self-reported indicators were then coded and categorized by two researchers. Overall, identified 17 indicators related to user behavior within the system and eight system event-based indicators. We also identified six physical reactions and eight emotions related to the need for explanation.

Our work contributes to the fields of requirements engineering and software engineering in several ways. First, the established indicators can be used to identify the need for explanations before the development using prototypes. Second, they can be used to automatically detect the explanation needs in existing applications using telemetry and usage data. Furthermore, these indicators can be used to detect the need for explanations at runtime in order to trigger explanations at the appropriate time to introduce explainability into adaptive personalized systems.

The remainder of the paper is structured as follows:

Section~\ref{sec:background} provides an overview of the background and related work. The study design is described in Section~\ref{sec:research}. Section~\ref{sec:results-indicators} and Section~\ref{sec:resultsII} present the study results, which are further analyzed and discussed in Section~\ref{sec:discussion}. Finally, Section~\ref{sec:conclusion} summarizes the conclusions and outlines directions for future work.


\section{Background and Related Work}
\label{sec:background}
\subsection{Explainability}
In the context of requirements engineering, explainability is considered a non-functional requirement and software quality aspect~\cite{chazette2020explainability,chazette2021exploring,deters2024x,kohl2019explainability}. In general terms, explainability describes the ability of a software to be explained to its stakeholders~\cite{kohl2019explainability}. 
Usually, explanations are implemented with the goal of improving the users' understanding of~\cite{adadi2018peeking,chazette2020explainability,jongeling2024towards} and experience with the software~\cite{deters2024x,droste2024explanations}.

Chazette et al~\cite{chazette2022explainable,chazette2021exploring} conducted a systematic literature review to research the relationship between explainability and other software quality aspects. Their findings indicate that while software-sided explanations can lead to positive effects on system aspects such as transparency, learnability and understandability, they might also lead to negative effects on other aspects, such as performance, security and development cost. Deters et al.~\cite{deters2024x} made similar observations in the context of user experience: For one, explanations may improve aspects like user efficiency and trust, by providing the necessary information to properly navigate and understand a system. However, the same aspects might also be negatively affected by overly complicated or complex explanations, confusing users and obstructing their experience rather than supporting them.

In the field of explainable artificial intelligence (XAI), the focus usually lies on reasoning decision making processes and justifying outputs, making the system more interpretable~\cite{adadi2018peeking,habiba2025ml}. However, explanations may also be used to make other system aspects more understandable. In the context of end-user privacy, Brunotte et al.~\cite{brunotte2023context,brunotte2023privacy} demonstrated that explanations are a suitable means to make software more transparent and trustworthy. This is also apparent in concepts such as explainable security~\cite{vigano2020explainable} and explainable hardware~\cite{speith2024explainability}. Furthermore, research in human-computer interaction has shown that explaining how to use a system can improve users' mental models of and proficiency with the system~\cite{kieras1984role,staggers1993mental}.

Droste et al.~\cite{droste2024explanations} developed a taxonomy for explainability needs in the context of everyday software systems. Their findings suggest that explanations are a suitable means to address a variety of different issues in software, not just \textit{unexpected system behavior}. In particular, the identified types of needs relate to explanations for \textit{interactions} between the user and the system, \textit{privacy and security} concerns, \textit{domain specific system elements} as well as \textit{user interfaces}. In a later work by Obaidi et al.~\cite{obaidi2025automating}, this taxonomy was successfully deployed in an industrial context, enabling the automation of explanation need management in mobile app reviews.

\subsection{Detection of Explanation Needs}
Identifying users' need for explanation is challenging due to various biases that are introduced during the elicitation process. Asking users openly about their need for explanations for a particular application overwhelms them and introduces a hypothetical bias~\cite{hypotheticalBiasPlott}. To mitigate this bias, Droste et al.~\cite{Droste2023personas} elicited explanation needs by suggesting types of explanations and asking users if they would like to receive them. Furthermore, they found that when asked if they want certain explanations in a software, users tend to agree regardless of if the explanations are actually needed and in spite of any negative effects that unnecessary explanations may have. They called this phenomenon the \textit{why-not mentality}.

To address these two biases, Deters et al.~\cite{Deters2023explanationsOnDemand} conducted a study in which they introduced an omnipresent explainer that provided explanations of various aspects at every point in the system. They then tracked which explanations were actually requested by their study participants. This method enabled users to take negative effects of explanations into account, counteracting the \textit{why-not mentality}. However, the approach is practically not feasible, as explanations for all parts of the system would have to be written in advance, introducing considerable time effort, especially for large systems. 

In the context of runtime indicators, Deters et al.~\cite{deters2024empaticaPreview} attempted to identify the need for explanations using biometric data. Metrics such as electrodermal activity and blood volume pulse are common indicators for measuring emotional states~\cite{girardi2022empatca,schmidt2019wearable}. Although some of their results were promising, the runtime analysis of biometric data is not a sufficient indicator at this stage to reliably detect the need for explanations~\cite{deters2024empaticaPreview}.

\subsection{Elicitation Methods}
The elicitation of requirements is a core activity in requirements engineering, as all subsequent activities build on its results~\cite{pacheco2018elicitationTechniquesSLR}. The integration of stakeholders plays a major role, since the software systems should be customized to their needs~\cite{mishra2008elicitationTechniquesCombination}. The most frequently used method for eliciting requirements are interviews~\cite{pacheco2018elicitationTechniquesSLR}. Interviews are a suitable method for getting a general overview of the requirements and for clarifying misunderstandings directly~\cite{pacheco2018elicitationTechniquesSLR}. According to Mishra et al.~\cite{mishra2008elicitationTechniquesCombination}, interviews are most effective when combined with workshops and iterative development to elicit and validate requirements. 

When it comes to requirements that stakeholders cannot express themselves or have yet to be determined, the use of prototypes has proven to be helpful~\cite{mannio2001elicitationPrototypesScenarios,pacheco2018elicitationTechniquesSLR,schneider1996protoytpesAsAssets,suranto2015protoypesElicitation}. Prototypes can serve as a communication aid and help to identify problems that were not previously considered, enabling the requirements to be refined and completed~\cite{pacheco2018elicitationTechniquesSLR}. Interactive high-fidelity prototypes are particularly useful for this purpose~\cite{suranto2015protoypesElicitation,schneider1996protoytpesAsAssets}, as they simulate real interactions and can therefore be used to determine quality requirements~\cite{mannio2001elicitationPrototypesScenarios}. The high-fidelity prototypes can be used to identify needs for explanation with the help of the indicators identified in this paper.

\section{Study Design}
\label{sec:research}
\subsection{Research Questions}

In this work, we take the first step towards creating a catalog that systematically organizes indicators that point to explanation needs. To achieve this, we pose the following research questions:
\begin{framed}
    \begin{enumerate}
        \item[RQ1] Which runtime indicators exist that signal a need for explanation?
        \item[RQ2] Which indicators point to which type of explanation needs?

    \end{enumerate}
\end{framed}
The first research question aims to collect which behaviors or emotions users report when they experience a need for explanation. These indicators should be recordable during runtime, i.e. when the user is using the system. The second research question then aims to establish a connection between these indicators and types of explanation needs. Doing this enables developers to choose the appropriate indicators if they want to examine a specific type of need for explanation.

\subsection{Survey Structure}
To gather indicators for explanation needs, we developed a survey that first collects recent explanation needs of the participants and then asks for their behavior and emotions associated with these needs. To this end, the survey is divided into three hierarchically structured parts, as shown in Figure~\ref{fig:structure}. 

\begin{figure}[h]
    \centering
    \includegraphics[width=.9\linewidth]{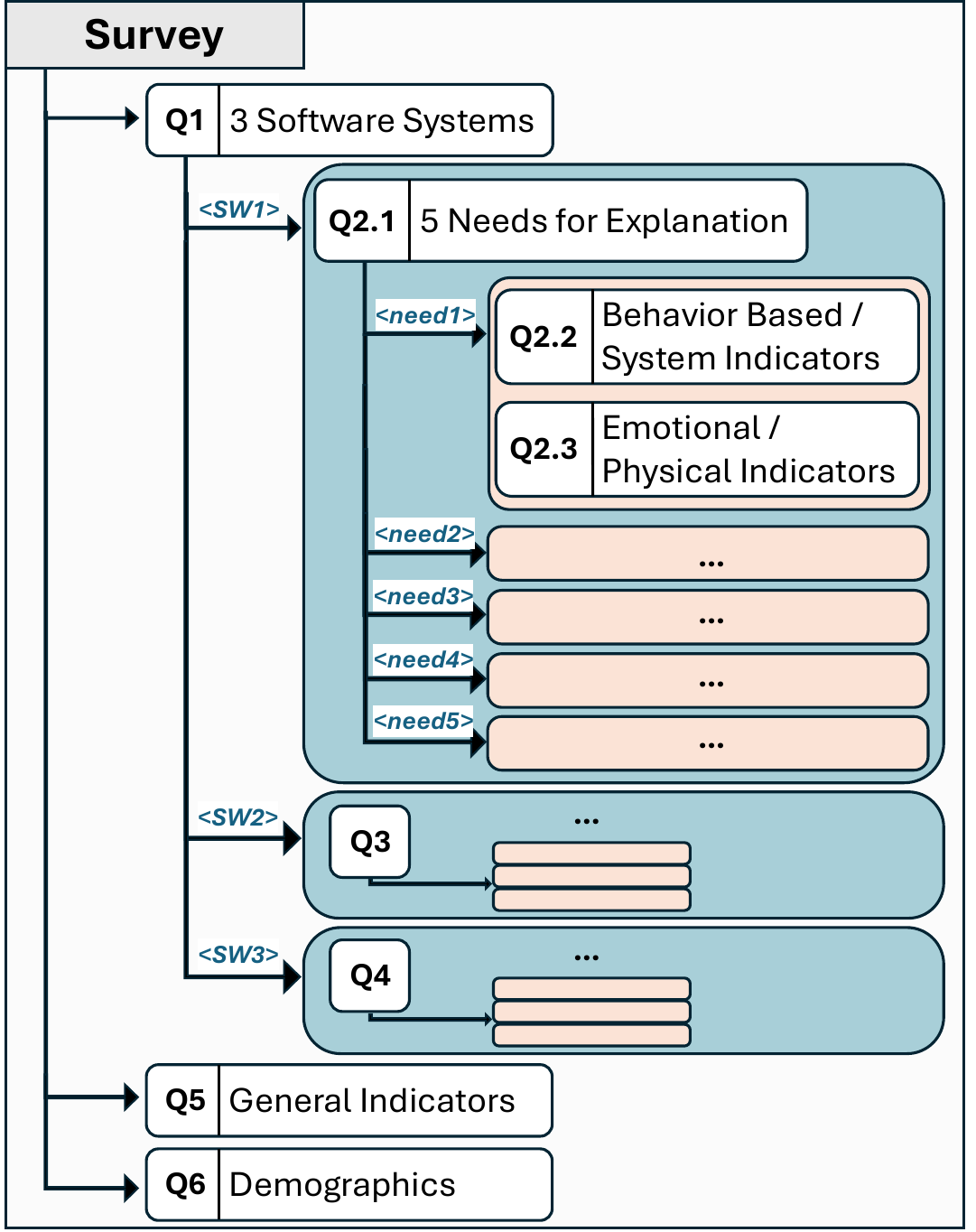}
    \caption{Structure of the survey}
    \label{fig:structure}
\end{figure}


To collect actual explanation needs, we first asked participants about what the three software systems were they had most recently used~(\textit{Q1}). This allowed the participants to set their own context for the survey, which mitigates hypothetical bias. For each of these three software systems, the participants could then report up to five explanation needs that arose during use~(blue boxes around \textit{Q2.1 - Q4.1} in Figure~\ref{fig:structure}). For each of these five explanation needs, we asked two follow-up questions to identify indicators for the specific explanation need (red boxes around Q2.2 and Q2.3 in Figure~\ref{fig:structure}). The first question was about how the participants behaved when an explanation need arose within the software~(\textit{Q2.2 - Q4.2}). The second question asked whether the explanation need had evoked physical or emotional reactions~(\textit{Q2.3 - Q4.3}). 

After collecting explanation needs and their associated indicators for the three software systems they mentioned, \textit{Q5} enabled the participants to name further indicators independently of the software systems mentioned in \textit{Q1}. This ensured that participants could report indicators that remained in their memory, even those not related to the last three software systems the participant used. 
Finally, demographic data such as age and field of work were collected~(\textit{Q6}).

We provided supporting examples and information for the questions to enable users to answer the questions as accurately as possible. For example, when eliciting explanation needs, it has proven to be effective to use a taxonomy that the participant can use to navigate~\cite{droste2024explanations, obaidi2025automating}. For this purpose, we used the taxonomy of Droste et al.~\cite{droste2024explanations} and gave an example for each of the types of explanation needs. This helped to stimulate the participants' memories and to point out different types of needs that may exist. The complete questionnaire is available in the supplementary material~\cite{supplemenatryMateiral}.


\subsection{Data Collection}
The survey was hosted on the LimeSurvey platform~\footnote{https://limesurvey.org} and was available throughout the last quarter of 2024. The survey was distributed via networking platforms and the institution's bulletin board. We deemed this kind of convenience sampling appropriate as the only requirement for participant was being of legal age (18 years).

Three types of data were collected. 
\begin{enumerate}
    \item Demographic data: We asked participants for their age, gender and field of work. This data is reported with descriptive analyses. 
    \item Software Systems: The participants reported three software systems in Q1, which consist of one or two words. This data was only used to enable the participants to set their own context and is not analyzed further in this paper. 
    \item Needs and Indicators: The participants reported their needs for explanation and indicators in the form of free text responses consisting of several words or sentences. This data is the focus of our analyses in this paper and was analyzed using coding procedures.
    \end{enumerate}

\subsection{Demographics}
A total of 66 participants completed the survey. This included 29 males, 37 females and no  participants identifying as diverse. The average age was 30.7 years (min: 19, max: 56, SD: 11.03). Among the participants, 32 people were employed, 15 were students, and another 15 were both studying and employed. Two people stated that they were unemployed and one person stated that they were a housewife. A large proportion of participants (36\%) worked in a technical field. 11\% of the participants stated that they were employed in the healthcare sector, 9\% in the skilled trades and 6\% in teaching. The remaining participants worked in other fields including retail, social services or logistics.

\subsection{Data Analysis}
For the analysis of the free-text responses from questions \textit{Q2} to \textit{Q5}, two authors coded the answers using the coding software MAXQDA~\footnote{www.maxqda.com}. This was done in 4 labeling rounds, whereby each round involved coding the answers to one question group (see Figure~\ref{fig:data-analysis}). The question group here refers to the repeated questions on different software systems or explanation needs. Note that, for example, questions \textit{Q2.1}, \textit{Q3.1} and \textit{Q4.1} are the same question on different software systems. To simplify this section, question groups are referenced with the first question, in this case, question \textit{Q2.1}. Table~\ref{tab:interrater} summarizes for each labeling round which questions were coded and which taxonomy was used in each labeling round.

\begin{figure}[htb]
    \centering
    \includegraphics[width=.95\linewidth]{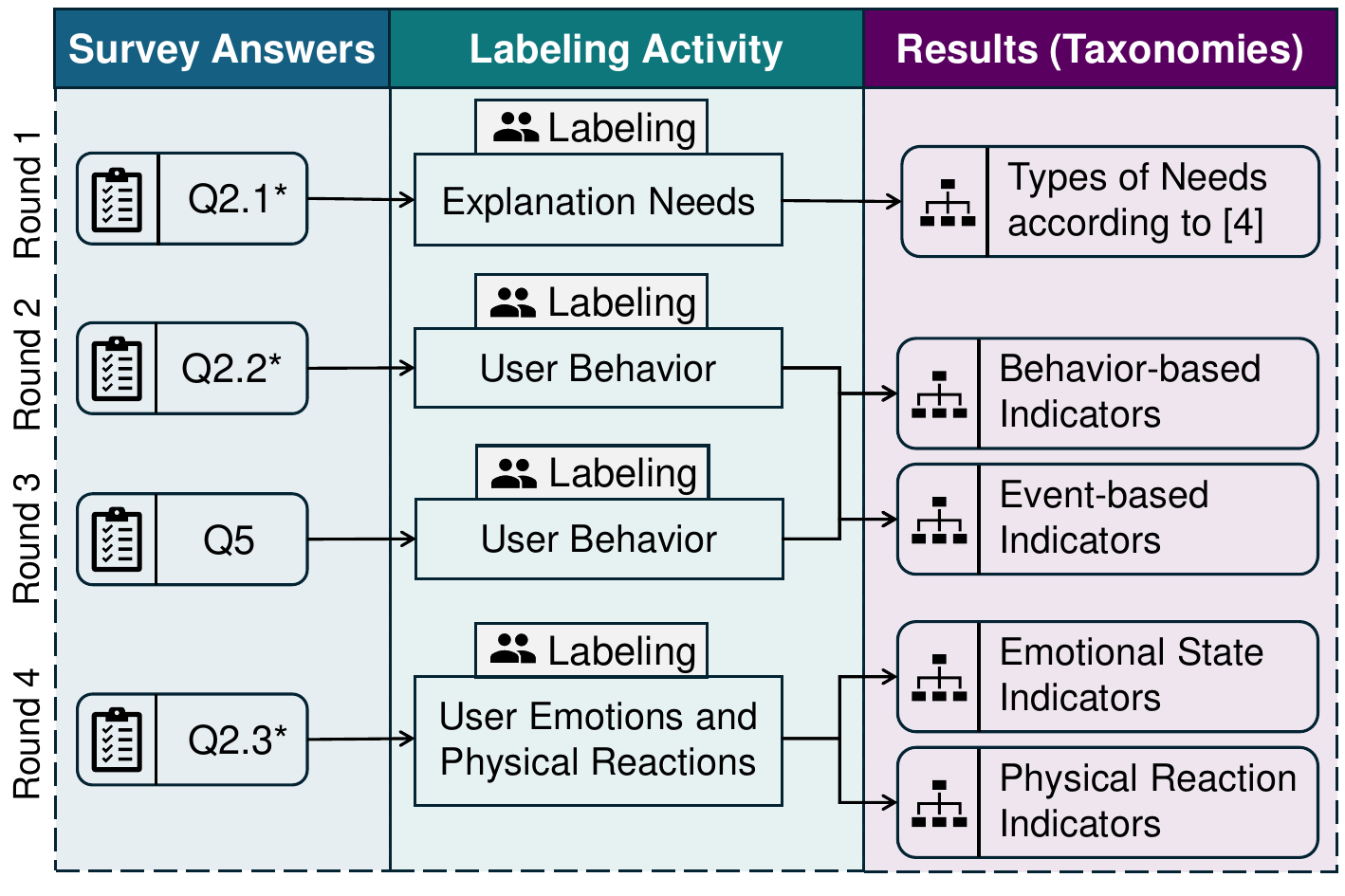}
    \footnotesize *Q2.x stands for questions Q2.x, Q3.x and Q4.x, which are the same questions for different software systems (see Figure~\ref{fig:structure}).
    \caption{Data analysis procedure}
    \label{fig:data-analysis}
\end{figure}

\textit{Round 1:} We used the explainability need taxonomy by Droste et al.~\cite{droste2024explanations} for the first labeling round for the answers of \textit{Q2.1}. The taxonomy divides explanatory needs into five categories, each with up to four subcategories. Two authors of this paper labeled based on the coding guidelines provided by Droste et al.~\cite{Droste2024SupMat} and subsequently resolved all conflicts.

\textit{Round 2+3:} In order to code question \textit{Q2.2}, we first needed to establish a taxonomy for behavior-based and event-based indicators. For this purpose, one author of the paper went through all the answers and created a preliminary taxonomy for labeling. This taxonomy regarding user behavior and system events was discussed and refined with another author. Based on this, we created coding guidelines\footnote{These guidelines are available in our supplementary material~\cite{supplemenatryMateiral}}. All answers to \textit{Q2.2} were then coded by two coders. The answers for question \textit{Q5} were categorized in the same taxonomy, again by two coders. If an answer did not fit into any existing category during coding, a new category was added to the taxonomy. 
Subsequently, all conflicts for \textit{Q2.2} and \textit{Q5} were resolved. 

\textit{Round 4:} For the responses to question \textit{Q2.3}, we developed a taxonomy regarding emotions and physical reactions. This taxonomy was again developed by one author and discussed and refined with another author. We also created coding guidelines~\cite{supplemenatryMateiral}. Again, two coders coded all answers of \textit{Q2.3} and for answers that did not fit into a category, a new one was added. We added two new categories regarding physical interaction during this labeling round. After that, all conflicts were resolved.

At the end of the four labeling rounds, all responses to the questions \textit{Q2.1}, \textit{Q2.2}, \textit{Q2.3} and \textit{Q5} had been categorized using the respective taxonomies.

\subsection{Interrater Agreement}
All conflicts that arose during the labeling process were successfully resolved, leaving no unresolved discrepancies. To calculate the interrater agreement, we used Brennan \& Prediger $\kappa$ (B \& P $\kappa$) \cite{brennan81kappa}, an adapted kappa from Cohen's $\kappa$ \cite{cohen60kappa}, which is suitable for two raters. Table~\ref{tab:interrater} presents the interrater agreement we achieved for each labeling round.

\begin{table}[]
\centering
\caption{Interrater agreement values during labeling}
\label{tab:interrater}
\begin{tabular}{@{}cp{1.4cm}p{3cm}l@{}}
\toprule
              & Questions        & Taxonomy                                             & B \& P $\kappa$  \\ \midrule
Round 1        & Q2.1, Q3.1, Q4.1 & Types of explanation needs~\cite{droste2024explanations}                          & \textbf{0.68}  \\ \hline
Round 2        & Q2.2, Q3.2, Q4.2 & Behavior-based and event-based indicators           & \textbf{0.61}  \\ \hline
Round 3        & Q5               & Behavior-based and event-based indicators           & \textbf{0.59}  \\ \hline
Round 4        & Q2.3, Q3.3, Q4.3 & Emotional state and physical reaction indicators    & \textbf{0.91}  \\ \bottomrule
\end{tabular}
\end{table}

According to Landis and Koch~\cite{landis1977measurement}, a  $\kappa$ value between $0.61 - 0.80$ represents a \textit{substantial} agreement. This means that we reached a \textit{substantial} agreement in round one and round two. In round three, only a \textit{moderate} agreement was reached, likely due to the more open-ended nature of responses to question 5. The answers to question 5 were therefore somewhat less clear than the answers to question \textit{Q2.2} (round two). In round four, we reached an \textit{almost perfect} agreement~\cite{landis1977measurement}.

\subsection{Data Availability Statement}
We share participants' responses in a pseundonomized form for questions \textit{Q1} - \textit{Q5}, which were labeled in our study. Responses that either indicated a refusal to answer or contained inappropriate information were excluded from the published dataset. We also provide the assigned codes of the corresponding taxonomy for the labeled answers. This allows for reproducibility and validation of our findings. However, to protect participants' privacy, we do not share responses to the demographic questions.

\section{Results - Indicators}
\label{sec:results-indicators}

While reviewing the participants' answers, we noticed that the participants' responses were not limited to the user's behavior in the system. In some cases, answers also described system-internal events that indicate a need for explanation regardless of the user's behavior. Therefore, we differentiated between behavior-based and event-based indicators, which resulted in two distinct taxonomies. In addition to the behaviors within the system, we also surveyed physical reactions and emotions, leading to two further taxonomies. Overall, we created four taxonomies containing different types of indicators, which are described below.

\subsection{Behavior-based Indicators}
The taxonomy for behavior-based indicators contains indicators based on how the user interacts with the system. Figure~\ref{fig:behavior-indicator} shows behavior-based indicators that were mentioned by participants in the survey. 

\begin{figure}[h]
    \centering
    \includegraphics[width=\linewidth]{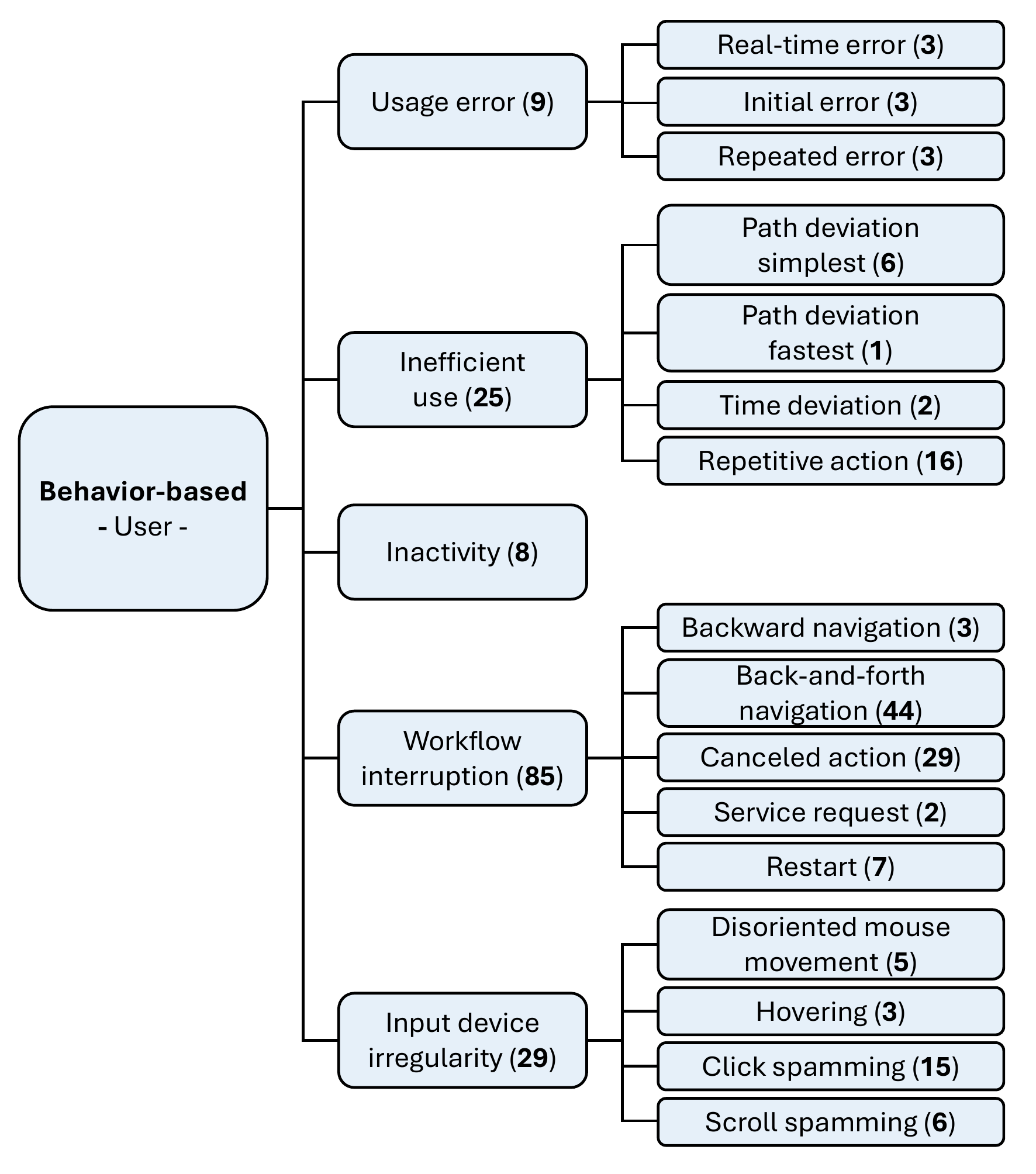}
    \footnotesize *The numbers in brackets reflect the number of times the indicator appeared in the participants' responses.
    \caption{Behavior-based Indicators}
    \label{fig:behavior-indicator}
\end{figure}

\textit{Usage Error.} This category of indicators relates to incorrect user input and was reported nine times. The responses from the participants revealed that the need for explanation arises at different error rates depending on the use case. In some cases, the need for explanation occurred in real time when an erroneous input was entered (\textbf{real-time error}). For example, if an incorrect syntax was used in an \textit{Excel} cell, participants wanted to receive an explanation immediately rather than after pressing enter. In other cases, errors cannot be recognized in real time, in which case an explanation is desired directly after the input (\textbf{initial error}). However, some explanations should not be immediately displayed when an error occurs for the first time. For example, if an incorrect password has been entered for the first time, the user may have simply mistyped it and does not need an explanation. However, if the user enters the password incorrectly several times, an explanation should be provided (\textbf{repeated error}).

\textit{Inefficient Use.} Indicators in this category relate to inefficient use of the system by the user. A total of 25 participants' responses included indicators from this category. For some of these indicators, it is necessary to first determine how a system should be used optimally. If a user deviates from the optimal path through a use case, this could be a sign of a need for explanation. Here we differentiate between the simplest path and the fastest path. The fastest path means that there are shortcuts or other abbreviations that can be used, but which may make the path more complicated (\textbf{path deviation - fastest}). The simplest path is about the ease of use of a coherent task (\textbf{path deviation - simplest}). For example, format suggestions in \textit{Word} are more convenient to use than creating a layout yourself. Another possibility is to recognize that a user needs an unusually long time for a task (\textbf{time deviation}). However, this requires the software to have strongly cohesive task strands that take approximately the same amount of time. The last indicator  in this category is \textbf{repetitive actions}, which was mentioned 16 times. For example, it could indicate that a user does not understand the search function if they keep scrolling through the search bar, repeatedly pressing the next button.

\textit{Inactivity.} The absence of user input for a longer period of time may also indicate a need for explanation (\textbf{Inactivity}). We have not identified any subcategories for this category.

\textit{Workflow Interruption.} Another indication of a need for explanation is the interruption of the workflow of a coherent task. This category is the most frequently reported with 85 participant responses. Navigating backwards suggests that the user wants to undo or change something (\textbf{backwards navigation}). This may indicate that there has been a misunderstanding. A similar principle applies when a task is canceled (\textbf{canceled action}). The task is also interrupted when the system is \textbf{restarted} with the user indicating that something is not working the way they want it to. Another type of workflow interruption is the \textbf{service request}. However, it should be noted that the user is already actively looking for help, so the need for explanation would be recognized rather late. \textbf{Back-and-forth navigation} is another type of workflow interruption which was reported 44 time. By navigating back and forth in the system (e.g., searching through the system settings) the user shows that they are looking for something and that they need guidance or specific information.

\textit{Input Device Irregularity.}
The unusual use of input devices could indicate a need for explanation, which was frequently reported by participants (29 times). Input devices include, for example, keyboards, mice and touchscreens. The first two indicators we extracted is the spamming of input. Scrolling up and down erratically indicates that the user is not really processing the content, but is instead expressing frustration or disorientation (\textbf{Scroll spamming)}. The same applies to frequent clicks on non-interactive objects (\textbf{Click spamming}). These clicks do not lead to intended actions, but are rather a sign of impatience or frustration. \textbf{Hovering} over elements for an extended period can also be a sign that the user needs more information, as they seem interested in the object but do not yet want to interact with it. \textbf{Disoriented mouse movements} across the screen, without targeted interaction, may also indicate a need for guidance.

\subsection{Event-based Indicators}
The taxonomy for event-based indicators contains indicators based on system-internal events that may cause a need for explanation regardless of user behavior. These events frequently trigger explanation needs for the majority of users.
\begin{figure}[h]
    \centering
    \includegraphics[width=\linewidth]{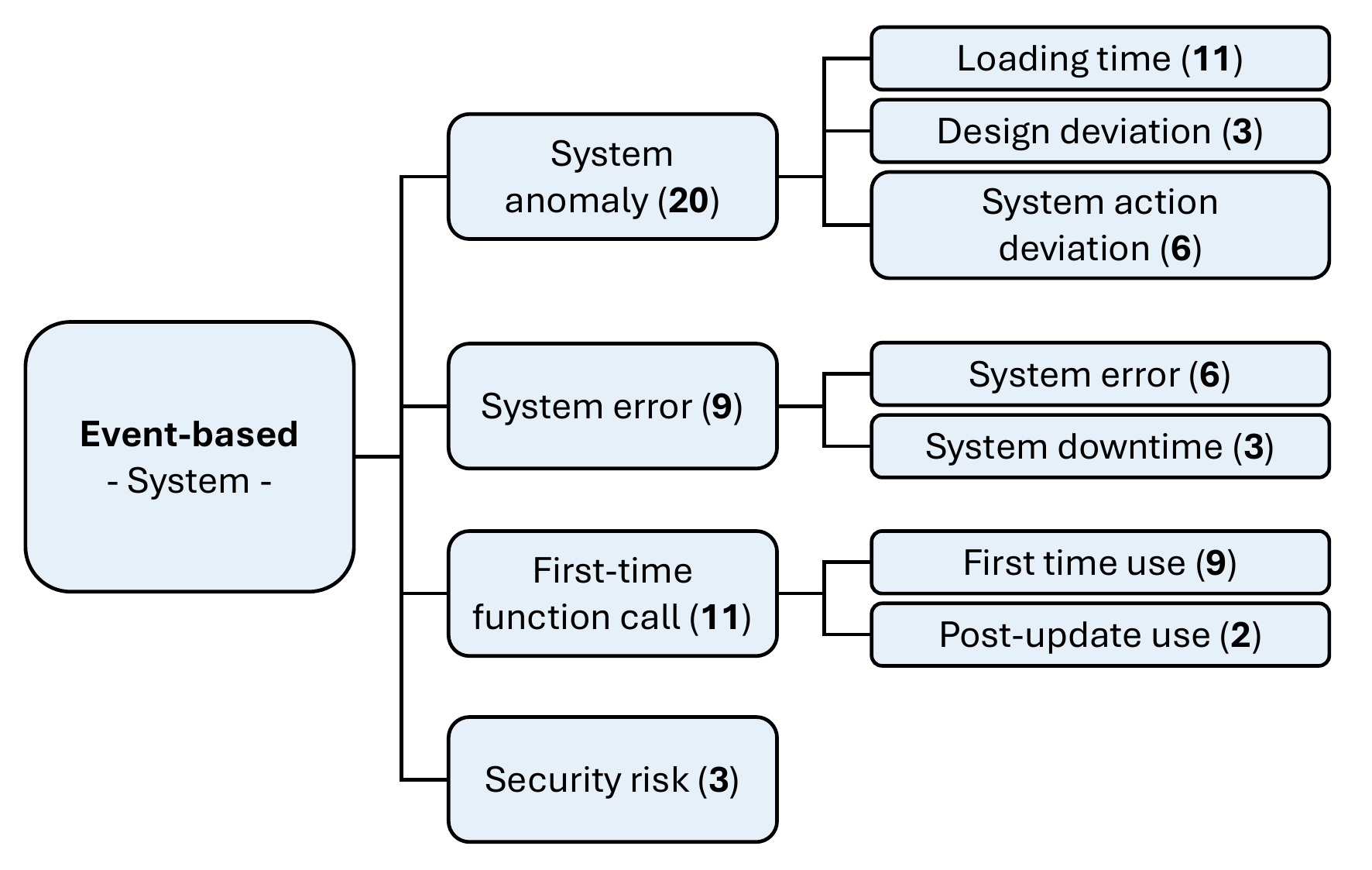}    \footnotesize *The numbers in brackets reflect the number of times the indicator appeared in the participants' responses.
    \caption{Event-based Indicators}
    \label{fig:event-indicator}
\end{figure}

\textit{System Anomaly.} 
This category includes unusual system behavior. Eleven participants stated that they wish to receive an explanation of what is happening during \textbf{loading times}. In addition, deviations from usual design principles or design changes after updates can also signal a need for explanation (\textbf{Design deviation}). For example, if the logout button is not in commonly used position at the top right, the user might become confused. Another event-based indicator is deviations from usual system reactions or responses (\textbf{System action deviation}). For example, if the navigation system suggests a route that differs from the usual route, users are likely to require an explanation.

\textit{System Error.}
System errors also create a need for explanation. This category occurred in 9 times in the participants' answers. System errors include simple failures of an action (\textbf{System error}), or even a complete system crash (\textbf{System downtime}). For these indicators, it is necessary that the system itself recognizes the error and provides an explanation of what went wrong.

\textit{First-time Function Call.}
When using (modified) functions for the first time, there is also an increased need for explanation, according to eleven survey responses. When functions are changed after an update, users may need an explanation of what has changed,  and why and how it can be used now (\textbf{Post-update use}). Independent of updates, complex functions may also require explanation when they are used for the first time (\textbf{First time use}). However, it is important to note that not every function should be explained the first time it is called, but only complex functions. In that context, whether or not a function should be considered complex depends on many factors and is thus subject to future research.

\textit{Security Risk.}
The system can also detect internally when an action poses \textbf{security risks}. Three participants stated that they want explanations in this case. We have not identified any subcategories for this category.

\subsection{Physical Reactions}
The indicators related to physical reactions do not involve direct interaction with the system, but require additional input devices for detection. For example, verbal expressions would need to be captured via a microphone These indicators may not yet ready for direct use, but only provide directions for future research. In the case of verbal expressions, for example, it would still have to be determined exactly which type of expression (e.g., sighing) indicates a need for explanation.
\begin{figure}[h]
    \centering
    \includegraphics[width=\linewidth]{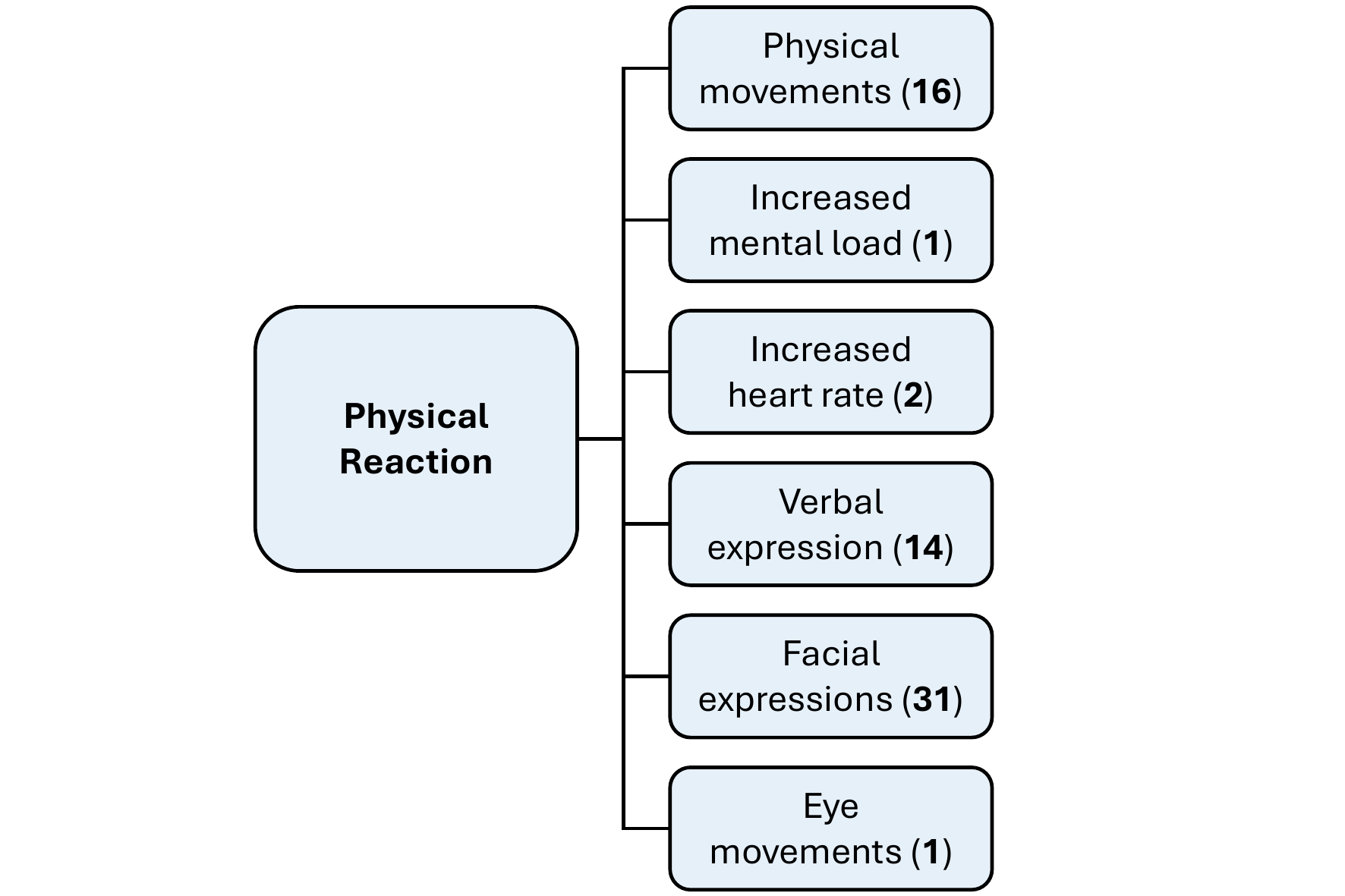}    \footnotesize *The numbers in brackets reflect the number of times the physical reaction was reported by the participants.
    \caption{Physical Reaction Indicators}
    \label{fig:physical-indicator}
\end{figure}

\textbf{Physical Movements} such as scratching the head or leaning towards the screen can indicate a need for explanation. One participant explicitly stated that they noticed they had an \textbf{increased mental load} when having a need for explanation. There are already several established methods for recording mental load, such as the pupil width~\cite{pfleging2016pupilDiameter} or other biometric values~\cite{kushki2011placementOfSensoresEDAandBVP, schmidt2019wearable}. These methods could therefore also be used to identify the need for explanation. Two participants also stated that an \textbf{increased heart rate} could indicate a need for explanation. 14 participants also reported \textbf{verbal expressions} such as sighing or annoyed groaning while having a need for explanation. Another physical indicator mentioned by the participants is \textbf{facial expressions}. For that, explicit features such as frowning or raised eyebrows can be examined. Furthermore, video recognition could be used to detect emotions from facial expressions. For this, the emotions listed below can be used. Finally, specific \textbf{eye movement} patterns could also provide information about the need for explanation, which can be recorded using eye trackers.

\subsection{Emotional State}
The indicators for emotional state work in a similar way to those for physical reactions. Again, there is no direct interaction with the system, but they would have to be captured with an additional device. There are already numerous studies on the emotional states of users in order to gain insights from them~\cite{girardi2022empatca, obaidi2025moodAndExpl}. We assume that certain emotions are triggered when there is a need for explanation, and that these emotions can be recognized through various methods. To advance research in this area, the taxonomy of emotional states offers initial insights into the emotions that users feel when they need explanations. Our taxonomy differentiates between negative and positive emotions to provide an overview.
\begin{figure}[h]
    \centering
    \includegraphics[width=\linewidth]{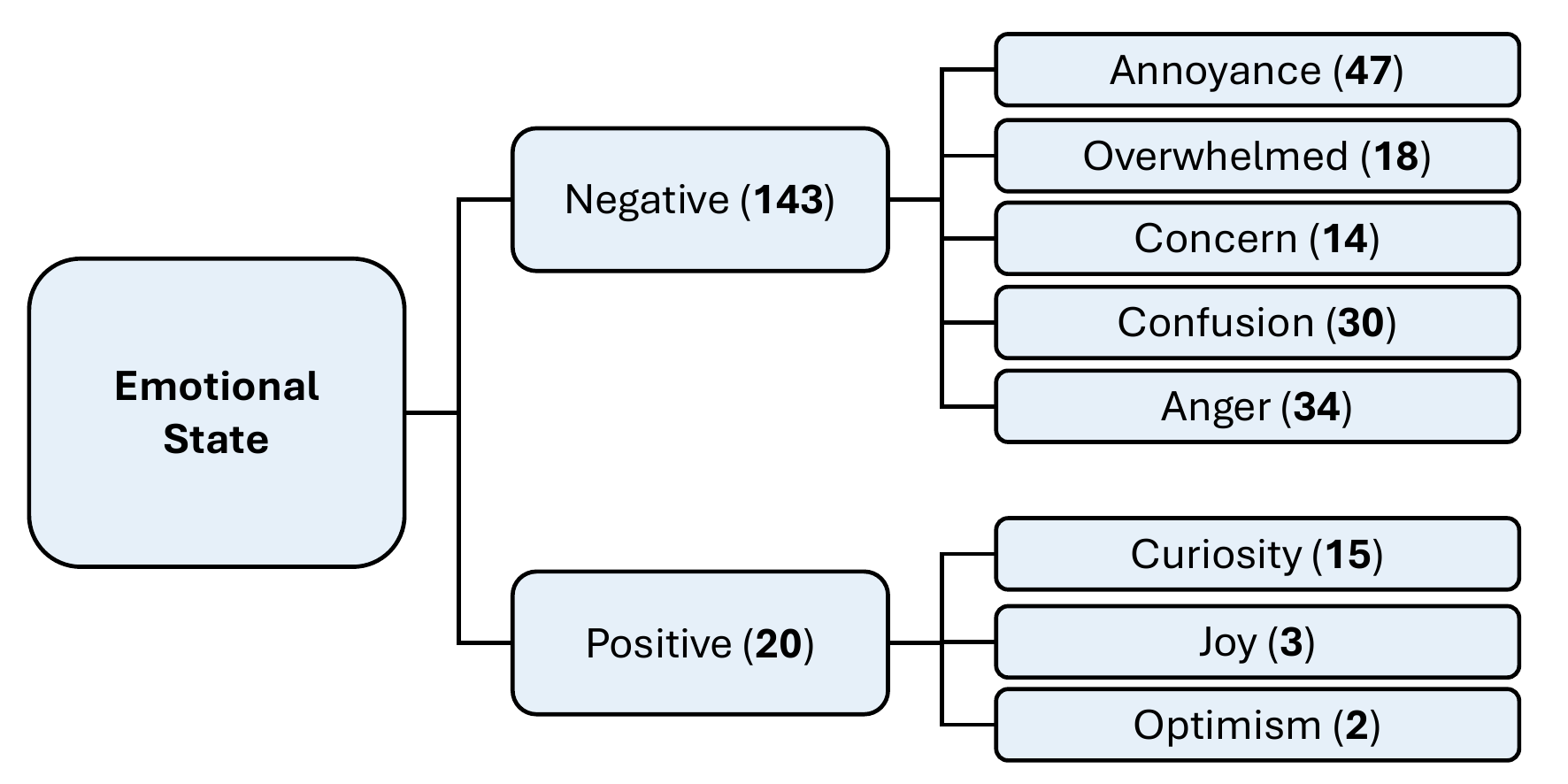}    \footnotesize *The numbers in brackets reflect the number of times the emotion was reported by the participants.
    \caption{Emotional State Indicators}
    \label{fig:emotional-indicator}
\end{figure}

\textit{Negative Emotions.}
143 responses from participants reported negative emotions related to the need for explanation. The emotion of \textbf{annoyance} includes impatience or displeasure, which was mentioned 47 times. The emotion of feeling \textbf{overwhelmed} encompasses feelings such as stress, tension and panic, which was reported 18 times. The emotion of \textbf{concern} was attributed when a user stated that they felt helpless or were afraid of doing something wrong. This was indicated 30 times. The emotion of \textbf{anger} was attributed when a participant stated that they were angry and aggressive, which occurred 34 times.

\textit{Positive Emotions.}
Some participants also mentioned positive emotions related to the expectation of an explanation. The participants mentioned 15 times that they felt \textbf{curious} while beeing interested in the information needed. In addition, three participants mentioned confidence, expressing \textbf{optimism} that the software would resolve their lack of understanding. Finally, two participants also indicated that they felt anticipation for an explanation (\textbf{joy}).

\section{Indicators and Types of Needs}
\label{sec:resultsII}
\subsection{User Behavior}
In the first labeling round, we categorized the types of explanation needs described by users. 
Table~\ref{tab:needsandindicators} shows how often each type of need was mentioned  by the participants. To illustrate the relationship between the identified types of needs and the indicators, Table~\ref{tab:needsandindicators} also lists the behaviors that were described when a particular type of need arose.

\begin{table}[tb]
\centering
\caption{Number of stated needs for explanation and associated behaviors.}
\label{tab:needsandindicators}
\begin{tabular}{l l l r}    
\toprule
\multicolumn{2}{c}{\textbf{Needs}} & \multicolumn{2}{c}{\textbf{Described Behaviors}} \\ 
Type of Need & Count & Indicator & Count \\ \midrule
\multirow{14}{*}{Interaction} & \multirow{14}{*}{99} & Back-and-forth navigation & 32 \\ 
 &  & Canceled action & 24 \\
 &  & Repetitive action & 6 \\
 &  & Inactivity & 4 \\
 &  & Click spamming & 4 \\
 &  & Scroll spamming & 3 \\
 &  & Disoriented mouse movement & 2 \\
 &  & Repeated error & 2 \\
 &  & Initial error & 1 \\
 &  & Path deviation (simplest) & 1 \\
 &  & Service request & 1 \\
 &  & Restart & 1 \\
 &  & Hovering & 1 \\
 &  & System action deviation & 1 \\ \midrule
\multirow{14}{*}{\begin{tabular}[l]{@{}l@{}}System \\ Behavior\end{tabular}} & \multirow{14}{*}{35} & Repetitive action & 5 \\
 &  & Restart & 5 \\
 &  & Back-and-forth navigation & 4 \\
 &  & Canceled action & 4 \\
 &  & Click spamming & 4 \\
 &  & Backward navigation & 2 \\
 &  & Disoriented mouse movement & 2 \\
 &  & Real-time error & 1 \\
 &  & Repeated error & 1 \\
 &  & Path deviation (simplest) & 1 \\
 &  & Hovering & 1 \\
 &  & Scroll spamming & 1 \\
 &  & System action deviation & 1 \\
 &  & System error & 1 \\ \midrule
Privacy and\\ Security & 11 & Back-and-forth navigation & 1 \\ \midrule
\multirow{3}{*}{\begin{tabular}[l]{@{}l@{}}Domain\\ Knowledge\end{tabular}} & \multirow{3}{*}{15} & Back-and-forth navigation & 2 \\
 &  & Real-time error & 1 \\
 &  & Inactivity & 1 \\ \midrule
\multirow{2}{*}{\begin{tabular}[l]{@{}l@{}}User \\ Interface\end{tabular}} & \multirow{2}{*}{3} & Back-and-forth navigation & 2 \\
 &  & Design deviation & 1 \\ \bottomrule

\end{tabular}

\end{table}

Table~\ref{tab:needsandindicators} shows that the type of need \textit{interaction} was most frequently mentioned by the participants, with 99 reported needs for explanation. Looking at the described behaviors, we see that in 82 cases, the participants described behavior that could be identified as a behavior-based indicator. Thus, in over 80\% of cases, the need for interaction explanation could have been identified on the basis of behavioral patterns.
Accordingly, approximately 90\% of the 35 reported needs for explanation regarding the \textit{system behavior} could have been identified through behavioral patterns of the user. 
 In contrast, only three needs were reported regarding the \textit{user interface}, of which two could have been predicted using behavioral patterns and one using an system event-based indicator.

With regard to domain knowledge, participants mentioned an explanation need 15 times. However, only a few behaviors (below 30\%) were mentioned that could have been identified to recognize the need for explanation. The same applies to \textit{privacy and security}, where only one out of eleven cases was associated with a recognizable behavior.

\begin{framed}
\noindent
    \textbf{Finding 1:} The explanation needs related to \textit{privacy and security} and \textit{domain knowledge} appear to be difficult to recognize based on user behavior and system events.
\end{framed}

The need for explanation regarding \textit{interaction} was most frequently reported in connection with \textbf{back-and-forth navigation} and \textbf{canceled action}. Both indicators fall under the category of workflow interruption and indicate that the task sequence was interrupted as the participants did not know how to interact next. Inefficient use was also mentioned quite frequently with the \textbf{repetitive action} indicator, which also makes sense, as participants are more likely to act inefficiently if they have difficulties with the interaction.

Explanation needs regarding \textit{system behavior} were often reported alongside the \textbf{repetitive action} pattern. This seems counter-intuitive at first, but becomes apparent as participants who do not understand why the system arrived at a certain result may have performed the action repeatedly in order to understand through trial and error. Action cancellations such as the patterns \textbf{restart}, \textbf{canceled action} and \textbf{backward navigation} were also reported frequently, as participants aborted actions when they were unsure about the inner workings of the system.

As mentioned earlier, almost no patterns were reported in connection with explanation needs for \textit{domain knowledge} and \textit{security and privacy}. Both are most likely to be recognized by search patterns, but due to the small number of reports, this statement cannot be generalized. The same applies to \textit{user interface}-related explanation needs.

Notably, the \textit{back-and-forth navigation} indicator seems to recognize a large proportion of explanation requirements of different types. In 30\% of cases where there was a need for explanation regarding interaction, this behavior pattern was reported. In the case of explanation needs regarding user interface, two out of three needs were reported with this pattern and in the case of explanation needs regarding system behavior, 10\% of the cases.

\begin{framed}
\noindent
    \textbf{Finding 2:} The indicator \textbf{back-and-forth navigation} seems to be well-suited for detecting explanation needs related to \textit{interaction}, \textit{system behavior} and \textit{user interface}.
\end{framed}

\subsection{Physical Reactions}

In addition to behavioral patterns, the participants could report physical reactions for each need they mentioned. Therefore, Table~\ref{tab:needsandreactionss} lists the same type of needs as table~\ref{tab:needsandindicators}, with the corresponding reactions instead of behavioral patterns.

\begin{table}[htb]
\centering
\caption{Stated needs for explanation and associated stated physical reactions.}
\label{tab:needsandreactionss}
\begin{tabular}{l l l r}    
\toprule
\multicolumn{2}{c}{\textbf{Needs}} & \multicolumn{2}{c}{\textbf{Described Physical Reactions}} \\ 
Type of Need & Count & Indicator & Count \\ \midrule
\multirow{6}{*}{Interaction} & \multirow{6}{*}{99} & Facial expressions & 17 \\ 
 &  & Physical movements & 9 \\
 &  & Verbal expression & 7 \\
 &  & Increased heart rate & 1 \\
 &  & Eye movements & 1 \\
 &  & Increased mental load  & 1 \\ \midrule
\multirow{3}{*}{\begin{tabular}[l]{@{}l@{}}System \\ Behavior\end{tabular}} & \multirow{3}{*}{35} & Facial expressions & 9 \\
 &  & Verbal expression & 4 \\
 &  & Physical movements & 3 \\ \midrule
\multirow{2}{*}{\begin{tabular}[l]{@{}l@{}}Privacy and\\ Security\end{tabular}} & \multirow{2}{*}{11} & Facial expressions & 1 \\
 &  & Increased heart rate & 1 \\ \midrule
\multirow{3}{*}{\begin{tabular}[l]{@{}l@{}}Domain\\ Knowledge\end{tabular}} & \multirow{3}{*}{15} & Physical movements & 4 \\
 &  & Facial expressions & 3 \\
 &  & Verbal expression & 2 \\ \midrule
\multirow{3}{*}{\begin{tabular}[l]{@{}l@{}}User \\ Interface\end{tabular}} & \multirow{3}{*}{3} & Facial expressions & 1 \\
 &  & Verbal expression & 1 \\
 &  & Physical movements & 1 \\ \bottomrule

\end{tabular}

\end{table}

Notably, 60\% of the domain knowledge-related needs evoked a physical reaction, while only a quarter evoked a specific behavioral pattern. This suggests that the need for explanation regarding domain knowledge can be better recognized through physical reactions than behavioral patterns. However, since the number of needs regarding domain knowledge is quite small, this statement cannot be generalized.

The change in facial expression was mentioned for all types of need. However, as this reaction can have many different manifestations, it is possible that different types of need for explanation cause different types of facial expressions.

\section{Discussion}
\label{sec:discussion}
\subsection{Interpretation of the results}
We identified four types of indicators for the need for explanation: behavioral patterns, system events, physical reactions and emotions. 
We derived 17 distinct behavior-based indicators from 156 described behavior patterns. More than half of the described behavior patterns fell into the category of workflow interruption. In particular, users frequently described how they navigated back and forth within the software, which shows that they were looking for something but couldn't find it. In addition, cancelling an action, such as aborting a purchase process, was also frequently described as an indicator of the need for explanation.

\begin{framed}
\noindent
    \textbf{Finding 3:} User behavior patterns indicating a \textbf{workflow interruption} are the most frequently reported indicators of a need for explanation. In particular, \textbf{back-and-forth navigation} and \textbf{canceled actions} are often described.
\end{framed}

While coding the reported behaviors, we noticed that some system events caused an increased need for explanation. Such system events were described 43 times by participants, from which we were able to derive eight indicators. These event-based indicators differ from the other indicators as they do not have to be collected with the help of prototypes and user studies, but can be collected statically from the system. For example, an explanation may be provided whenever a system crashes. The most frequently reported indicator was system behavior that deviates from the normal behavior of the system like unusually long loading times and unusual system responses. Note that these indicators are only a preliminary collection, and further research is needed into which other events might exist and how often they actually cause a need for explanation.

\begin{framed}
\noindent
    \textbf{Finding 4:} Some system events inherently create a need for explanation. The most frequently reported event that triggered an explanation need was the system behaving different from usual (\textbf{system anomaly}).
\end{framed}

The third type of indicators are physical reactions. The participants reported physical reactions linked to the need for explanation 65 times. From these 65 descriptions we were able to extract 6 distinct physical reactions. The most frequently mentioned physical reaction was a change in facial expression. We were not able to determine what type of facial expression indicated a need for explanation, which needs to be investigated in further studies.

\begin{framed}
\noindent
    \textbf{Finding 5:} A change in \textbf{facial expression} seems to be a promising physical indicator of the need for an explanation.
\end{framed}

As one might intuitively expect, the need for explanations triggers negative emotions more often than positive ones. \textit{Annoyance} was mentioned most frequently. Interestingly, negative emotions were triggered quite frequent by the need for explanations. In over 60\% of the cases where a need for explanation arose, the participants explicitly reported a negative emotion. This highlights the importance of addressing users' need for explanation, as negative emotions, which could impair user retention, should be prevented by all means.
\begin{framed}
\noindent
    \textbf{Finding 6:} The need for explanation often triggers negative emotions, with \textbf{Annoyance} being the most prevalent.
\end{framed}

\subsection{Implications}

\subsubsection{Limitations of the Found Indicators}
While our work identified many useful indicators based on user reports, it should be noted that the taxonomies described in Section~\ref{sec:results-indicators} are not yet a complete catalog of indicators. Firstly, it can be assumed that more indicators exist, especially indicators that users perform unconsciously. This would require, for example, exploratory experiments in which users are being observed.
Secondly, participant reports do not provide evidence of how well the identified indicators function in practice. This also needs to be investigated in practice. Some indicators as reported are not yet directly applicable. In particular, the reported physical reactions only provide rough directions, but not yet ready-to-use methods. This would require future research into how exactly the physical reactions can be recorded and interpreted. The user behavior patterns are the indicators that were most accurately described by the participants and are therefore the most ready-to-use indicators. However, some details still need to be determined, such as how many clicks in a row count as click spamming.

Since the details and exact procedures for such indicators depend heavily on the specific use case, a universal catalog of ready-to-use indicators is unrealistic. Instead, this catalog should serve as a framework that requirements engineers and developers can build upon to identify suitable indicators. These indicators function as modular components that must be adapted to the specific use case. The exact details of their implementation must therefore be determined by engineers themselves. That being said, general guidelines could be beneficial—for instance, defining click spamming as five to ten clicks within a short period. 

\subsubsection{Choosing Indicators}
As described in Section~\ref{sec:results-indicators}, some indicators were reported more frequently than others. It can be assumed that indicators that were mentioned more frequently by participants are more likely to be suitable for many use cases. For example, indicators that were only reported by one participant may not be generalizable to other users and therefore not feasible. Nevertheless, we decided to include all reported indicators in order not to erroneously exclude any indicators. This work is intended as a first step towards an applicable catalog in which we aim to collect initial approaches to indicators as broadly as possible. 

For the practical application of the indicators, we recommend choosing the indicators that were reported more frequently. However, most important is the fit to the use case, which depends partly on the type of explanation needed, but also on system-specific constraints.

\subsubsection{Intended Use of the Indicators}
The identified indicators are intended to objectively recognize a need for explanation. We envision four primary use cases:

\textit{Requirements Engineering with Prototypes.} First, the indicators can be used in the requirements engineering process using high-fidelity prototypes. Users will be advised to use the prototypes as naturally as possible while recording their interactions and reactions. The interactions and reactions can then be analyzed using the indicators to identify areas in the system where a need for explanation arose. Follow-up interviews could provide additional insights.

\textit{Post-Deployment Analysis.} The second option involves already published software systems that are in use. In these systems, telemetry and usage data could be recorded and analyzed retrospectively. In particular, behavior-based indicators can be used here to identify behavior patterns that may indicate problems.

\textit{Real-time Explanation Triggers.} The third possible deployment option is explanation triggers. Here, the indicators could be used to recognize the need for explanations from users at runtime in order to provide explanations with an appropriate timing. To this end, the indicators would have to be included in the source code in order to evaluate them automatically during runtime.

\textit{Advancing Explainability Research.} The fourth area of application is research. Explainability is a well-researched topic, but it is difficult to objectively measure whether or not the need for explanation has been met. These indicators could be used to advance research in the area of explainability by providing more objective measures than questionnaires and interviews.

\subsubsection{Explanation Needs $\neq$ Explainability Requirements}
The identified indicators recognize the need for explanations regarding software systems. It should be noted that the need for explanation does not always have to be solved with explanations. For example, it might be possible, that the need for explanation regarding interaction can also be solved with improved usability. In many cases, an explanation is the simplest or most effective solution, but it should be decided on a case-by-case basis what the best solution is for this problem. Thus, we want to emphasize that the indicators are not exclusively detectors for explainability requirements, but rather point to general problems that may be solvable with the help of explainability.

\subsection{Threats to Validity}
We report the threats to the validity of our research according to Wohlin et al.~\cite{wohlin2012experimentation}:

\paragraph{Construct Validity.}
Due to the design of the study, we could only identify and analyze self-reported behaviors. Unconscious behaviors or imperceptible physical reactions could not be detected with our study design. However, self-reported behaviors are the only way to collect a broad range of initial approaches for indicators, as experiments could not be carried out with as many different software systems and participants. Building upon this initial work, future studies can be carried out to identify unconscious behaviors and thus enhance the catalog.

\paragraph{Internal Validity.}
The accuracy of the reported behaviors and reactions relies on participants’ ability to recall past experiences. Some participants may have recalled detailed and precise interactions, while others may have forgotten or simplified their experiences. This variability introduces inconsistency in the dataset. Furthermore, participants' answers may be subject to availability bias or recall bias. To counteract these, we allowed participants to indicate the last software systems they used, so that they could set a context for themselves. The process of coding user feedback is always subjective in nature. Therefore, it is possible that different coders would have categorized the data in a different manner. To counteract this, we created detailed coding guidelines to make the process as objective and transparent as possible. 

\paragraph{External Validity.}
Our study included a variety of software applications, but it is unclear whether the identified indicators apply equally across different types of software. For example, explanation needs in productivity software (e.g., word processors, spreadsheets) may differ significantly from those in entertainment applications (e.g., video streaming platforms, games). Furthermore, with a sample size of 66 participants, it is likely that we did not capture all existing behaviors indicative of a need for explanation. However, as  our sample included a diverse range of participants, we are confident that the most prominent indicators were identified in our study.

\paragraph{Conclusion Validity.}
While our study identifies correlations between certain behaviors and explanation needs, it does not establish causation. Just because a participant navigated back and forth in a system does not necessarily mean they needed an explanation. As we did not perform a hands-on validation of the indicators, we cannot draw any conclusions about the actual applicability of the indicators. However, as this work is intended to be a starting point for a catalog of indicators, and thus mainly to motivate further research, proving the actual applicability is not yet necessary. 

The subjectivity of our coding procedures also affects the conclusion validity. We achieved at least substantial agreement in all but one coding round (which barely missed the threshold). This suggests a solid connection between the data set and the assigned codes. As such, different researchers would likely have come to the similar conclusions.

\section{Conclusion and Future Work}
\label{sec:conclusion}
Identifying needs for explanation is a challenging task, as it is subject to many biases, such as hypothetical bias or the \textit{why-not mentality}. To overcome these biases, we aim at creating a catalog of indicators that signal a need for explanation.
As a first step in this direction, we conducted a survey in which participants reported behaviors and reactions associated with the need for explanation. We coded these responses in four labeling rounds and established four types of indicators.

We identified 17 indicators based on behavior patterns, eight indicators based on specific system events, and six indicators based on physical reactions. We also identified which emotions are associated with a need for explanation, thus highlighting the importance of addressing these needs. We also established a link between types of explanation needs and indicators, simplifying the selection of appropriate indicators. However, the indicators should be considered preliminary as they have not yet been validated, which is planned in future research.
With applicable indicators, we contribute to many areas of software engineering. Indicators can be used in regular elicitation processes, can help to automatically identify the need for explanations from usage data, can help to provide explanations at the right time, and can be used in explainability research.

In future research, we plan to expand and refine the catalog. To expand the catalog, we intend to conduct experiments that record unconscious behavior. To this end, we plan to confront participants with explanatory needs in software and observe their behavior. To improve the catalog, we envision two steps. First, we plan to test the indicators for their applicability and to evaluate them using various criteria such as precision, recall and implementation effort. Second, we plan to establish practical guidelines for applying the indicators to improve their applicability.


\bibliographystyle{IEEEtran}
\bibliography{references}

\end{document}